\documentclass[pra,aps,10pt,twocolumn,showpacs,amsmath,amssymb]{revtex4-1}
\usepackage{graphicx}

\begin{document}
\title{Plasmon effects in photoemission}
\author{J. Z. Kami\'{n}ski\thanks{e-mail: jkam@fuw.edu.pl}}
\affiliation{Institute of Theoretical Physics, Faculty of Physics,
University of Warsaw, Pasteura 5, 02-093 Warszawa, Poland
}
\date{\today}
%\maketitle

\begin{abstract}
We develop the concept of scattering matrix and we use it to perform 
stable numerical calculations of photo-emission from nano-tips. Electrons move 
in an external space and time dependent nonperturbative electric field. 
We apply our algorithm for different strengths and spatial configurations of the field.   

\end{abstract}
\maketitle

\section{Introduction}

The aim of this paper is to investigate some particular quantum processes taking place in an arbitrary space-dependent scalar potential and a time- and space-dependent vector potential. Vector potential is periodic in time and describes a laser field. Its space-dependence results from the interaction of the laser field with electrons in solids. Such conditions are met for example in semiconductor nanostructures \cite{Hamilton2015,FG1989,K1993} (like quantum wires or wells), photoemission from a metal tip \cite{Kim2008,Kruger2011}, carbon nanotubes or graphene \cite{Zhou2011,KDB2009,HR2006} or in surface physics \cite{Mahmood2016,Wang2013,Wang2015,FAGAS2009,SMAMK2008,FGM2007,DKF2006}. To make our presentation as clear as possible we shall restrict ourselves to the one-space-dimensional case, although extension of the presented method to systems of higher dimensionality is possible (see, e.g. \cite{FKS2005}). We shall apply our method to investigation of the photo-emission process.

This paper is organized as follows. In Sec. II the most general solution of the Schr\"odinger equation is introduced. The transfer-matrix method and matching conditions are analyzed in Sec. III, 
whereas reflection and transition probabilities are introduced in Sec. IV. These probabilities must sum up to 1, which puts a very strong check for the accuracy of numerical calculations. The most important part of this paper, i.e. the concept of the scattering-matrix method, is discussed in the next section, where it is shown why the scattering-matrix algorithm has to be introduced, instead of a much simpler transfer-matrix algorithm. Numerical illustrations of the applicability of this algorithm are presented in Sec. VI, and are followed by short conclusions.

In our numerical illustrations we use atomic units, if otherwise stated.

\section{Solution of the Schr\"odinger equation}

Let us start with one-dimensional Schr\"odinger equation of the form \cite{LL92},
\begin{align}
\mathrm{i}\partial_{t}\psi(x,t)=&\Bigl [ \frac{1}{2} \Bigl
(\frac{1}{\mathrm{i}}\partial_{x}-eA(x,t)\Bigr
)\frac{1}{m(x)}\Bigl (\frac{1}{\mathrm{i}}
\partial_{x}-eA(x,t)\Bigr )\nonumber \\
+&V(x)\Bigr ]
\psi(x,t).
\label{sec_1_1}
\end{align}
Space-dependent mass $m(x)$, scalar potential $V(x)$ and
 vector potential $A(x,t)$ are spatially constant in finite
intervals. Their values in any interval $(x_{i-1},x_{i})$
will be denoted as $m_{i}$, $V_{i}$ and
$A_{i}(t)$. We require also that the function $A(x,t)$ is periodic
in time, that is
\begin{equation}
A(x,t+T)=A(x,t),
\label{sec_1_2}
\end{equation}
where $T=2\pi/\omega$ and $\omega$ is the frequency of the
oscillating in time electric field.
Defining in a standard way the probability density
$\rho(x,t)$,
\begin{equation}
\rho(x,t)={\arrowvert\psi(x,t)\arrowvert}^{2},
\label{sec_1_3}
\end{equation}
and the probability current $j(x,t)$,
\begin{eqnarray}
j(x,t)&=&\frac{1}{2}{\psi^{\mathrm{*}}(x,t)}\frac{1}{m(x)} \Bigl
(\frac{1}{\mathrm{i}}\partial_{x}-eA(x,t)\Bigr )\psi(x,t) \\
&+&\frac{1}{2}\psi(x,t)\frac{1}{m(x)} \Bigl [ \Bigl
(\frac{1}{\mathrm{i}}\partial_{x}-eA(x,t)\Bigr ) \psi(x,t)\Bigr
]^{\mathrm{*}}, \nonumber \label{sec_1_4}
\end{eqnarray}
we show  using Eq. (\ref{sec_1_1}) that the conservation 
of probability condition  is satisfied. Indeed,
assuming the above definitions,
we get the continuity equation,
\begin{equation}
\partial_t \rho(x,t)+\partial_x j(x,t)=0.
\label{sec_1_4a}
\end{equation}
Space dependence of mass in Eq. (\ref{sec_1_1}) 
forces one to impose non-standard continuity
conditions on any solution of this equation. It is now
the wavefunction $\psi(x,t)$ and the quantity
\begin{equation}
\frac{1}{m(x)}\Bigl (\frac{1}{\mathrm{i}}\partial_{x}-eA(x,t)\Bigr
)\psi(x,t) \label{sec_1_5}
\end{equation}
that have  to be continuous at points of discontinuity of mass
$m(x)$ and both potentials $V(x)$ and $A(x,t)$ \cite{LL92,KE99,ML01,SK03}.
Before passing to a general solution $\psi(x,t)$ of Eq. (\ref{sec_1_1})
in any given interval $(x_{i-1},x_{i})$, which we shall denote as
$\psi_{i}(x,t)$, let us note that due to time periodicity of the
Hamiltonian, $\psi_{i}(x,t)$ can be chosen such that the Floquet
condition,
\begin{equation}
\psi_{i}(x,t+T)={\mathrm{e}}^{-\mathrm{i}ET}\psi_{i}(x,t),
\label{sec_1_6}
\end{equation}
is satisfied,
where $E$ is the so-called quasienergy.
A general solution $\psi_{i}(x,t)$ of Eq. (\ref{sec_1_1}) in any
interval $(x_{i-1}, x_{i})$ takes then  the following form \cite{K90,FK97},
\begin{eqnarray}
\psi_{i}(x,t)&=&\sum_{M=-\infty}^{\infty} \exp{\bigl
(-\mathrm{i}(E+M\omega)t\bigr )} \sum_{\sigma=\pm}^{}\label{sec_1_7} \\
 &\times&\sum_{N=-\infty}^{\infty}C_{iN}^{\sigma}\mathcal{B}_{M-N}(\sigma
p_{iN})\exp{(\mathrm{i}\sigma p_{iN}x)},
\nonumber
\end{eqnarray}
where $C_{iN}^{\sigma}$ are arbitrary complex numbers to be determined and
\begin{equation}
p_{iN}=\sqrt{2m_{i}(E+N\omega-V_{i}-U_{i})},
\label{sec_1_8}
\end{equation}
with $U_i=e^2\langle A_i^2(t)\rangle/2m_i$ being the ponderomotive energy,
where $\langle A_i^2(t)\rangle$ means the time-average of
$A_i^2(t)$ over the laser-field oscillation.
Components for which $p_{iN}$ are purely imaginary are called
closed channels. These channels are not observed for a particle in
initial or  final states, but they have to be taken into account
in order to satisfy the unitary condition of the time evolution.
In a general case, the $\mathcal{B}_{M-N}(\sigma p_{iN})$ functions
are components of the following Fourier expansion,
\begin{equation}
\exp{\bigl (\mathrm{i}\Phi_{iN}^{\sigma}(t)\bigr
)}=\sum_{M=-\infty}^{\infty}\exp{(-\mathrm{i}M\omega
t)}\mathcal{B}_{M-N}(\sigma p_{iN})
\end{equation}
 provided that the vector potential $A(x,t)$
is periodic in time.
Functions $\Phi_{iN}^{\sigma}(t)$ are defined as follows:
\begin{equation}
\Phi_{iN}^{\sigma}(t)=
\int_{0}^{t}\Bigl [
\frac{\sigma e }{m_{i}}A_{i}(t)p_{iN}-\frac{e^2}{2m_{i}}\bigl (
A_{i}^2(t)-\langle A_{i}^2(t)\rangle\bigr )
\Bigr ]\mathrm{d} t.
\end{equation}
It is easily seen from the above equation that the 
$\mathcal{B}_{M-N}(\sigma p_{iN})$ functions
depend on the form of the vector potential $A(x,t)$, 
that is on the laser field applied.

\section{Matching conditions and transfer matrix}

Continuity conditions discussed above and applied to a general
solution (\ref{sec_1_7}) of the Schr\"odinger equation
(\ref{sec_1_1}) lead to an infinite chain of equations connecting
constants $C_{iN}^{\sigma}$ in the neighboring domains. These
matching conditions can be written in the matrix form,
\begin{equation}
B(i-1,x_{i-1})C_{i-1}=B(i,x_{i-1})C_{i},
\label{sec_2_1}
\end{equation}
where $C_{iN}^{\pm}=[C_{i}^{\pm}]_{N}$ are the components of the
columns $C_{i}^{\pm}$. The matrices $B(i,x)$ and $C_{i}$ are
defined as follows,
\begin{equation}
B(i,x) =
\left( \begin{array}{cc}
B^{+}(i,x) & B^{-}(i,x) \\
{B'}^{+}(i,x) & {B'}^{-}(i,x)
\end{array} \right),
\,
C_{i} =
\left( \begin{array}{c}
C_{i}^{+} \\
C_{i}^{-}
\end{array} \right).
\label{sec_2_2}
\end{equation}
The elements of $B(i,x)$ can be computed in the following way.

For an arbitrary function $A(x,t)$, periodic in time with the period $T$ we have
\begin{equation}
A(x,t)=\sum_{n=-\infty}^{\infty}b_{n}(x)\exp{(-\mathrm{i}n\omega t)},
\label{sec_2_2_a}
\end{equation}
where $\omega=2\pi/T$.
In the interval $(x_{i-1}, x_{i})$  coefficients $b_{n}(x)$ assume 
constant values, which we shall denote as
$b_{i,n}$.
Using the condition of the continuity of the wavefunction $\psi_{i}(x,t)$ 
at the point $x_{i-1}$,
we compute the elements of the matrices
$B^{+}$ and $B^{-}$,
\begin{eqnarray}
{B}^{\pm}(i,x)_{M,N} & = & \mathcal{B}_{M-N} (\pm p_{i,N})
\exp{(\pm \mathrm{i}p_{i,N}x)}.
\label{sec_2_5}
\end{eqnarray}
On the other hand elements of the $B'$ matrix  can be evaluated by
substituting a general solution (\ref{sec_1_7}) 
to the expression (\ref{sec_1_5})
and applying the continuity condition to it at
$x_{i-1}$.  After some algebraic manipulations we 
obtain finally the expression for the $B'$-matrices,
\begin{eqnarray}
{B'}^{\pm}(i,x)_{M,N} & = & \pm \frac{1}{m_{i}}\mathcal{B}_{M-N}
(p_{i,N})p_{i,N}\exp{(\pm \mathrm{i}p_{i,N}x)} \nonumber \\
& - & \frac{1}{m_{i}}\sum_{n=-\infty}^{\infty}eb_{i,n}
\mathcal{B}_{M-N-n} (\pm p_{i,N})\nonumber \\
&\times&\exp{(\pm \mathrm{i}p_{i,N}x)},
\label{sec_2_10}
\end{eqnarray}
and a set of equations for vectors $C_{i}$,
\begin{equation}
C_{i} = B_{i} C_{i-1},
\label{sec_2_11}
\end{equation}
where
\begin{equation}
B_{i} = [B(i,x_{i-1})]^{-1} B(i-1,x_{i-1}).
\label{sec_2_12}
\end{equation}
These relations allow to connect a solution in a
given domain $x_{i-1}<x<x_{i}$ with an analogous solution in any
other domain $x_{j-1}<x<x_{j}$,
\begin{equation}
C_{j} = B_{j} B_{j-1} ,\ldots, B_{i+1} C_{i} = \mathcal{T}_{ji} C_{i},
\label{sec_2_7a}
\end{equation}
where $\mathcal{T}_{ji}$ is the so-called transfer 
matrix \cite{TE73,K90,ML01}.

\section{Reflection and transition probabilities}

It is clear now that on the basis of Eq.(\ref{sec_2_7a}) we can
connect solutions in the boundary domains $(-\infty,x_{0})$ and
$(x_{L-1},\infty)$. Values of mass $m(x)$, scalar
potential $V(x)$ and vector potential $A(x,t)$ in these
domains will be denoted as $m_{0}$, $V_{0}$, $A_{0}(t)$ and
$m_{L}$, $V_{L}$, $A_{L}(t)$, respectively. We can then write down
 solutions of (\ref{sec_1_1}) for each of these domains. These
solutions represent incident ($\psi_{\mathrm{inc}}$), reflected
($\psi_{\mathrm{ref}}$) and transmitted ($\psi_{\mathrm{tr}}$) waves, and
take the following form,
\begin{eqnarray}
\psi_{\mathrm{inc}}(x,t)&=&\sum_{M=-\infty}^{\infty}
   \exp{(-\mathrm{i}Et)}\exp{(-\mathrm{i}M\omega t)}\nonumber \\
   &\times&\mathcal{B}_{M}(p_{0})
   \exp(\mathrm{i}p_{0}x),
\label{unit_1}
\end{eqnarray}
\begin{eqnarray}
\psi_{\mathrm{ref}}(x,t)&=&\sum_{N,M=-\infty}^{\infty}
C_{0,N}^{-}\exp{(-\mathrm{i}Et)}
   \exp{(-\mathrm{i}M\omega
   t)}\nonumber \\
   &\times&\mathcal{B}_{M-N}(-p_{N})\exp{(-\mathrm{i}p_{N}x)},
\label{unit_2}
\end{eqnarray}
\begin{eqnarray}
\psi_{\mathrm{tr}}(x,t)&=&\sum_{M=-\infty}^{\infty}C_{L,N}^{+}
\exp{(-\mathrm{i}Et)}
 \exp{(-\mathrm{i}M\omega t)}\nonumber \\
 &\times&\mathcal{B}_{M-N}(q_{N}) \exp{(\mathrm{i}q_{N}x)},
\label{unit_3}
\end{eqnarray}
where
\begin{eqnarray}
p_{N}&=&\sqrt{2m_{0}(E+N\omega-V_{0}-U_{0})}, \nonumber \\
q_{N}&=&\sqrt{2m_{L}(E+N\omega-V_{L}-U_{L})}.
\label{unit_4}
\end{eqnarray}
Constants $C_{0,N}^{-}$ and $C_{L,N}^{+}$ will be denoted from now
on as $\textsc{R}_{N}$ and $\textsc{T}_{N}$, respectively. Using 
continuity conditions for functions defined above, we get the
probability conservation equation for reflection and transition
amplitudes, $\textsc{R}_{N}$ and $\textsc{T}_{N}$,
\begin{equation}
   \sum_{N\geqslant N_{\mathrm{ref}}}^{}
   \frac{p_{N}}{p_{0}}{\arrowvert \textsc{R}_{N}\arrowvert}^{2}
   +\sum_{N\geqslant N_{\mathrm{tr}}}^{}\frac{m_{0}q_{N}}{m_{L}p_{0}}
   {\arrowvert \textsc{T}_{N}\arrowvert}^{2}=1,
\label{unit_5}
\end{equation}
where summations are over such $N$ for which $p_N$ and $q_N$ are real,
i.e., over the open channels.
This equation permits us to interpret
\begin{equation}
P_{\mathrm{R}}(N)=\frac{p_{N}}{p_{0}}{\arrowvert 
\mathrm{R}_{N}\arrowvert}^{2}
\label{unit_6}
\end{equation}
and
\begin{equation}
P_{\mathrm{T}}(N)=\frac{m_{0}q_{N}}{m_{L}p_{0}}{\arrowvert \mathrm{T}_{N}
\arrowvert}^{2}
\label{unit_7}
\end{equation}
as reflection and transition probabilities for a tunneling
process in which absorption ($N>0$) or emission ($N<0$) of
energy $N\omega$ by electrons occurred \cite{K90,SK03}. In case of a
monochromatic laser field this process can be interpreted as 
absorption or emission of $N$ photons from the laser field.

The unitary condition (\ref{unit_5}) can be also interpreted as the conservation of
electric charge. To this end, let us define the quantities proportional to the
density of electric currents,
\begin{eqnarray}
J_{\mathrm{inc}}&=&\frac{p_0}{m_0}, \label{currinc}\\
J_{\mathrm{ref}}&=&\sum_{N\geqslant N_{\mathrm{ref}}}
\frac{p_N}{m_0}{\arrowvert \mathrm{R}_{N}\arrowvert}^{2}, \label{currref}\\
J_{\mathrm{tr}}&=&\sum_{N\geqslant N_{\mathrm{tr}}}
\frac{q_N}{m_L}{\arrowvert \mathrm{T}_{N}\arrowvert}^{2}. \label{currtr}
\end{eqnarray}
Then Eq. (\ref{unit_5}) adopts the form of the first Kirchhoff low,
\begin{equation}
J_{\mathrm{inc}}=J_{\mathrm{ref}}+J_{\mathrm{tr}}. \label{current}
\end{equation}

Using (\ref{sec_2_7a}) we can calculate constants $C_{0,N}^{-}=\mathrm{R}_{N}$
and $C_{L,N}^{+}=\mathrm{T}_{N}$ appearing in equations (\ref{unit_1}) -
(\ref{unit_3}). Indeed, since
\begin{equation}
C_{L}=\mathcal{T}C_{0},
\label{unit_8}
\end{equation}
where transfer matrix $\mathcal{T}=\mathcal{T}_{L0}$, and
because $\mathcal{T}$, $C_{0}$
and $C_{L}$ adopt the following block forms,
\begin{equation}
\mathcal{T}=
\left( \begin{array}{cc}
\mathcal{T}^{++} & \mathcal{T}^{+-} \\
\mathcal{T}^{-+} & \mathcal{T}^{--}
\end{array} \right),
C_{0} =
\left( \begin{array}{c}
C_{0}^{+} \\
\mathrm{R}
\end{array} \right),
C_{L}=
\left( \begin{array}{c}
\mathrm{T} \\
0
\end{array} \right),
\label{unit_9}
\end{equation}
we arrive at
\begin{eqnarray}
\textsc{T}&=&\mathcal{T}^{++}C_{0}^{+}+\mathcal{T}^{+-}\textsc{R}, \nonumber \\
0&=&\mathcal{T}^{-+}C_{0}^{+}+\mathcal{T}^{--}\textsc{R}, \label{unit_11}
\end{eqnarray}
where $\textsc{R}$ and $\textsc{T}$ denote the columns of $\textsc{R}_{N}$ i
$\textsc{T}_{N}$, and $[C_{0}^{+}]_{N}=\delta_{0,N}$. Thus, after some algebraic
manipulations, we have,
\begin{eqnarray}
\textsc{R}&=&-(\mathcal{T}^{--})^{-1}\mathcal{T}^{-+}C_{0}^{+}. \nonumber \\
\textsc{T}&=&\bigl (\mathcal{T}^{++}-\mathcal{T}^{+-}
    (\mathcal{T}^{--})^{-1}\mathcal{T}^{-+}\bigr )C_{0}^{+},
\label{unit_12}
\end{eqnarray}
which allows us to determine the quantities $\textsc{R}_{N}$ and
$\textsc{T}_{N}$ for a given transfer matrix $\mathcal{T}$. For
open channels, these quantities are the amplitudes of 
reflection ($\textsc{R}_{N}$) and transition ($\textsc{T}_{N}$)
probabilities, from which one can compute reflection and
transition probabilities using equations (\ref{unit_6}) and
(\ref{unit_7}). 

\section{The scattering matrix}

We note from equations (\ref{sec_2_5}) and (\ref{sec_2_10}) that each of 
the $B_{i}$ matrices that constitute the
transfer matrix $\mathcal{T}_{ji}$
contain elements $\exp(\pm \mathrm{i}p_{i,N}x_{i})$ that depend on the 
$x_{i}$ coordinates at which
the discontinuities appear. For closed channels, that is when the 
$p_{i,N}$ momenta are purely imaginary,
these numbers are real and may assume arbitrary values, very large 
or very small, depending again on the $x_{i}$ coordinates. Number of the
$B_{i}$ matrices is equal to the number of discontinuity points, that is it 
depends on how we divide the
space into short intervals in order to make our potential tractable by 
our algorithm. It may
therefore turn out that in order to compute the transfer matrix 
$\mathcal{T}_{ji}$, we have to multiply a large number
of the $B_{i}$ matrices, each containing both very small and very 
large numbers. It is clear that such a procedure is numerically unstable.
We have to find a way to modify our method of 
calculations in order to compute the elements of each $B_{i}$
matrix at the same point $x=0$ independently of where the `real' 
$x_{i}$ is. This would eliminate "dangerous"
$\exp(\pm \mathrm{i}p_{i,N}x_{i})$ elements (turning them to $1$),
however at the cost of appearing somewhere else.
We shall see later that these `left-overs' of the
shift into $x=0$ appear only as differences $x_{i+1}-x_{i}$ 
and therefore do not cause any harmful side-effects.
We shall see now that such a modification is possible and the price we 
pay for it is worth the effort.

It follows from Eq. (\ref{sec_2_7a}) that in the neighboring
domains, $(x_{i-2},x_{i-1})$ and $(x_{i-1},x_{i})$, we have,
\begin{equation}
C_{i} = \mathcal{T}_{i,i-1} C_{i-1}.
\label{sec_3_1}
\end{equation}
Although the elements of the transfer matrix 
$\mathcal{T}_{i,i-1}$ have been computed
from the continuity conditions at point $x_{i-1}$, one
can compute them at any other point, for example $x=0$. To this end, let us
notice what follows from the form of the solution (\ref{sec_1_7}). 
Translation of the system by a certain distance $\delta$ along the $x$-axis
causes only multiplication of each member of the sum over $N$ in
(\ref{sec_1_7}) by a constant  $\exp{(\mathrm{i}\sigma p_{iN}\delta)}$. These
constants can be included in coefficients $C_{iN}^{\sigma}$. In this way
we get a new set of constants which we shall denote as
$\tilde{C}_{iN}^{\sigma}$,
\begin{equation}
\tilde{C}_{iN}^{\sigma}=\exp{(\mathrm{i}\sigma p_{iN}\delta)} C_{iN}^{\sigma}.
\label{sec_3_2}
\end{equation}
We shall interpret these constants as coefficients in solution
(\ref{sec_1_7}),
given by the continuity conditions at  point 
$x_{i-1}-\delta$. Eq. (\ref{sec_3_2}) written
in the matrix form becomes,
\begin{equation}
\tilde{C}_{i}=\mathcal{P}_{i}(\delta)C_{i}
\label{sec_3_3}
\end{equation}
where
\begin{equation}
\mathcal{P}_{i}(\delta)=
\left( \begin{array}{cc}
 P_{i}^{+}(\delta) & 0 \\
 0 & P_{i}^{-}(\delta)
\end{array} \right),
\label{sec_3_4b}
\end{equation}
and
\begin{equation}
C_{i} =
\left( \begin{array}{c}
C_{i}^{+} \\
C_{i}^{-}
\end{array} \right),
\tilde{C}_{i} =
\left( \begin{array}{c}
\tilde{C}_{i}^{+} \\
\tilde{C}_{i}^{-}
\end{array} \right).
\label{sec_3_4}
\end{equation}
In the equation above $P_{i}^{\sigma}(\delta)$ is a diagonal matrix,
\begin{equation}
[P_{i}^{\sigma}(\delta)]_{NN'}=\delta_{NN'}\exp{(\mathrm{i}\sigma p_{iN}\delta)},
\label{sec_3_5}
\end{equation}
whereas $C_{i}^{\pm}$ and $\tilde{C}_{i}^{\pm}$ are the columns of the constants
 $C_{iN}^{\pm}$ and $\tilde{C}_{iN}^{\pm}$ respectively, that is
$[C_{i}^{\pm}]_{N}=C_{iN}^{\pm}$ and
$[\tilde{C}_{i}^{\pm}]_{N}=C_{iN}^{\pm}$.
It follows from the form of the matrix $\mathcal{P}_{i}(\delta)$ that the
following relations are satisfied:
\begin{equation}
\mathcal{P}_{i}^{-1}(\delta)=\mathcal{P}_{i}(-\delta),
\label{sec_3_6}
\end{equation}
\begin{equation}
\mathcal{P}_{i}(\delta_{1})\mathcal{P}_{i}(\delta_{2})
=\mathcal{P}_{i}(\delta_{1}+\delta_{2}).
\label{sec_3_7}
\end{equation}

Let us notice also that  translation of the system defined above modifies
the transfer matrix $\mathcal{T}_{i,i-1}$. We have
\begin{eqnarray}
\mathcal{P}_{i}^{-1}\tilde{C}_{i}=C_{i}
&=&\mathcal{T}_{i,i-1}C_{i-1}\nonumber \\
&=&\mathcal{T}_{i,i-1}\mathcal{P}_{i-1}^{-1}(\delta)
\mathcal{P}_{i-1}(\delta)C_{i-1},
\label{sec_3_8}
\end{eqnarray}
thus
\begin{eqnarray}
\tilde{C}_{i}=\mathcal{P}_{i}(\delta)\mathcal{T}_{i,i-1}\mathcal{P}_{i-1}^{-1}
(\delta)\tilde{C}_{i-1},
\label{sec_3_9}
\end{eqnarray}
and we can write it down as
\begin{equation}
\tilde{C}_{i}=\tilde{\mathcal{T}}_{i,i-1}\tilde{C}_{i-1},
\label{sec_3_10}
\end{equation}
where
\begin{equation}
\tilde{\mathcal{T}}_{i,i-1}=\mathcal{P}_{i}(\delta)\mathcal{T}_{i,i-1}
\mathcal{P}_{i-1}^{-1}(\delta).
\label{sec_3_11}
\end{equation}
Matrix elements denoted with the tilde symbol refer to the translated
system.  Using the method defined above and the relation (\ref{sec_2_7a}), we
can connect now the solution in the domain $(-\infty,x_{0})$ with the solution
in any other domain $(x_{i-1},x_{i})$. In this way the elements of the transfer
matrix, which have been computed until now at the points of discontinuity
$x_{0}\ldots x_{i-1}$, are computed now each time at the same point $x=0$. Let us
illustrate this method for a special case of $i=3$
\begin{eqnarray}
C_{3}&=&\mathcal{T}_{3,2}\mathcal{T}_{2,1}\mathcal{T}_{1,0}C_{1}=
\mathcal{P}_{3}^{-1}(x_{2})\mathcal{T}_{3,2}^{0}\mathcal{P}_{2}(x_{2})
\mathcal{P}_{2}^{-1}(x_{1})\nonumber \\
&\times&\mathcal{T}_{2,1}^{0}
\mathcal{P}_{1}(x_{1})\mathcal{P}_{1}^{-1}(x_{0})\mathcal{T}_{1,0}^{0}
\mathcal{P}_{0}(x_{0})C_{0} \nonumber \\
&=&\mathcal{P}_{3}^{-1}(x_{2})\mathcal{T}_{3,2}^{0}
\mathcal{P}_{2}(x_{2}-x_{1})\nonumber \\
&\times&\mathcal{T}_{2,1}^{0}\mathcal{P}_{1}(x_{1}-x_{0})
\mathcal{T}_{1,0}^{0}\mathcal{P}_{0}(x_{0})C_{0}.
\label{sec_3_12}
\end{eqnarray}
Equation (\ref{sec_3_12}) connects  constants $C_{0}$ and $C_{3}$ using the
matrices
$\mathcal{T}_{j,j-1}^{0}$ all computed at $x=0$ independently of $j$, and
 diagonal matrices $\mathcal{P}_{j}(\delta_{j})$, given by the relations
(\ref{sec_3_4b}) and (\ref{sec_3_5}), where $\delta_{j}=(x_{j}-x_{j-1})$.
Edge matrices $\mathcal{P}_{0}(x_{0})$ and $\mathcal{P}_{3}^{-1}(x_{2})$
in the equation (\ref{sec_3_12}) can be omitted 
while computing the transmission and
reflection probability amplitudes since their only role is to multiply the
amplitudes by  phase quotients which disappear while computing the
probabilities. Although these matrices lead to significant modifications of the
closed channels in the domains of $x<x_{0}$ and $x>x_{3}$ in this particular
case, these channels do not influence the reflection and transition 
amplitudes. Transmission and reflection probabilities 
can thus be computed  using a modified transfer matrix,
\begin{equation}
\mathcal{T}_{3,0}^{0}=\mathcal{T}_{3,2}^{0}\mathcal{P}_{2}(x_{2}-x_{1})
\mathcal{T}_{2,1}^{0}\mathcal{P}_{1}(x_{1}-x_{0})\mathcal{T}_{1,0}^{0}.
\label{sec_3_13}
\end{equation}
The matrices $\mathcal{T}_{i,i-1}^{0}$ are equal to the matrices $B_i$
in Eq. (\ref{sec_2_12}) calculated however for $x_{i-1}=0$.
This fact speeds up numerical calculations since now matrix 
$B(i,x=0)$ in Eq. (\ref{sec_2_12}) have to be inverted only once.
Further on we shall omit the superscript $0$ in $\mathcal{T}$ and the tilde over $C$
in order to simplify notation. 

The method presented above is still numerically unstable. The reason for this
instability lies in the existence of large numerical values of elements of
the  $\mathcal{P}_{i}^{-}(\delta)$ matrix for imaginary momenta $p_{iN}$.
In other words, for
\begin{eqnarray}
C_{i}&=&
\left( \begin{array}{c}
C_{i}^{+} \\
C_{i}^{-}
\end{array} \right)=
\mathcal{T}_{i,i-1}C_{i-1}\nonumber \\
&=&
\left( \begin{array}{cc}
\mathcal{T}_{i,i-1}^{++}  & \mathcal{T}_{i,i-1}^{+-}\\
\mathcal{T}_{i,i-1}^{-+}  & \mathcal{T}_{i,i-1}^{--}
\end{array} \right)
\left( \begin{array}{c}
C_{i-1}^{+} \\
C_{i-1}^{-}
\end{array} \right),
\label{sec_3_14}
\end{eqnarray}
the source of  numerical instabilities are matrix elements
$\mathcal{T}_{i,i-1}^{--}$ that contain large numbers.
There is however a chance  for improving the stability,
if only its reverse will be used, 
$\bigl (\mathcal{T}_{i,i-1}^{--}\bigr )^{-1}$.
It appears that it is possible provided 
that in our numerical algorithm only the
so-called scattering matrix will be applied. For this reason
we will show below how to compute the scattering matrix,
$\mathcal{S}_{j,i}$, using only elements of
the transfer matrix, $\mathcal{T}_{j,i}$.
For the transfer matrix $\mathcal{T}_{j,i}$ we have,
\begin{equation}
\left( \begin{array}{c}
C_{j}^{+} \\
C_{j}^{-}
\end{array} \right)=
\left( \begin{array}{cc}
\mathcal{T}_{j,i}^{++}  & \mathcal{T}_{j,i}^{+-}\\
\mathcal{T}_{j,i}^{-+}  & \mathcal{T}_{j,i}^{--}
\end{array} \right)
\left( \begin{array}{c}
C_{i}^{+} \\
C_{i}^{-}
\end{array} \right).
\label{sec_3_15}
\end{equation}
Thus,
\begin{eqnarray}
C_{j}^{+}&=&\mathcal{T}_{j,i}^{++}C_{i}^{+}+\mathcal{T}_{j,i}^{+-}C_{i}^{-},
\nonumber \\
C_{j}^{-}&=&\mathcal{T}_{j,i}^{-+}C_{i}^{+}+\mathcal{T}_{j,i}^{--}C_{i}^{-}.
\label{sec_3_16}
\end{eqnarray}
On the basis of (\ref{sec_3_16}) we now want to compute the elements of the
$\mathcal{S}_{j,i}$ matrix. This matrix 
is supposed to connect the coefficients
$C_{i}^{\pm}$ and $C_{j}^{\pm}$ in the following 
way,
\begin{eqnarray}
\left( \begin{array}{c}
C_{i}^{-} \\
C_{j}^{+}
\end{array} \right)=
\left( \begin{array}{cc}
\mathcal{S}_{j,i}^{++}  & \mathcal{S}_{j,i}^{+-}\\
\mathcal{S}_{j,i}^{-+}  & \mathcal{S}_{j,i}^{--}
\end{array} \right)
\left( \begin{array}{c}
C_{i}^{+} \\
C_{j}^{-}
\end{array} \right).
\label{sec_3_17}
\end{eqnarray}
Using the set of linear equations (\ref{sec_3_16}), we easily compute the
coefficients
$C_{i}^{-}$ and $C_{j}^{+}$ on the left-hand side of equation
(\ref{sec_3_17}),
as functions of the coefficients $C_{j}^{-}$ and $C_{i}^{+}$.
We get then the following relations,
\begin{eqnarray}
C_{i}^{-}&=&(\mathcal{T}_{j,i}^{--})^{-1}
(C_{j}^{-}-\mathcal{T}_{j,i}^{-+}C_{i}^{+}),
\nonumber \\
C_{j}^{+}&=&\bigl ( \mathcal{T}_{j,i}^{++}-\mathcal{T}_{j,i}^{+-}
(\mathcal{T}_{j,i}^{--})^{-1}
\mathcal{T}_{j,i}^{-+}\bigr )C_{i}^{+}\nonumber \\
&+&\mathcal{T}_{j,i}^{+-}
(\mathcal{T}_{j,i}^{--})^{-1}C_{j}^{-}.
\label{sec_3_18}
\end{eqnarray}
Finally we compute the elements of the matrix $\mathcal{S}_{j,i}$,
\begin{eqnarray}
\mathcal{S}_{j,i}^{++}&=&-(\mathcal{T}_{j,i}^{--})^{-1}
\mathcal{T}_{j,i}^{-+}, \nonumber \\
\mathcal{S}_{j,i}^{+-}&=&(\mathcal{T}_{j,i}^{--})^{-1}, \nonumber \\
\mathcal{S}_{j,i}^{-+}&=&\bigl ( \mathcal{T}_{j,i}^{++}-\mathcal{T}_{j,i}^{+-}
(\mathcal{T}_{j,i}^{--})^{-1}\mathcal{T}_{j,i}^{-+}\bigr ), \nonumber \\
\mathcal{S}_{j,i}^{--}&=&\mathcal{T}_{j,i}^{+-}
(\mathcal{T}_{j,i}^{--})^{-1}.
\label{sec_3_19}
\end{eqnarray}
As expected, the matrix $\mathcal{S}_{j,i}$ contains only  numerically
stable elements $(\mathcal{T}_{j,i}^{--})^{-1}$.

It follows from Eq. (\ref{sec_2_7a}) that the transfer matrix
$\mathcal{T}_{j,i}$ can be written as the product of two transfer matrices,
$\mathcal{T}_{j,k}$ and $\mathcal{T}_{k,i}$ ($i<k<j$),
\begin{eqnarray}
\mathcal{T}_{j,i}=\mathcal{T}_{j,k}\mathcal{T}_{k,i},
\end{eqnarray}
where matrices  $\mathcal{T}_{j,k}$ and $\mathcal{T}_{k,i}$ 
are defined as follows,
\begin{eqnarray}
C_{k}&=&\mathcal{T}_{k,i}C_{i},\nonumber \\
C_{j}&=&\mathcal{T}_{j,k}C_{k}.
\end{eqnarray}
Applying the method presented above, for each of the transfer matrices
$\mathcal{T}_{j,k}$ and $\mathcal{T}_{k,i}$
we can construct a scattering matrix,
$\mathcal{S}_{j,k}$ and $\mathcal{S}_{k,i}$ respectively.
Elements of the scattering matrix $\mathcal{S}_{j,i}$ can
be  computed using only elements of the matrices
$\mathcal{S}_{j,k}$ and $\mathcal{S}_{k,i}$. Using the 
notation above, we obtain the following
expressions for the elements of the $\mathcal{S}_{j,i}$ matrix,
\begin{align}
\mathcal{S}_{j,i}^{++}=&\mathcal{S}_{k,i}^{++}+\mathcal{S}_{k,i}^{+-}
   (1-\mathcal{S}_{j,k}^{++}\mathcal{S}_{k,i}^{--})^{-1}
   \mathcal{S}_{j,k}^{++}\mathcal{S}_{k,i}^{-+},\nonumber \\
\mathcal{S}_{j,i}^{+-}=&\mathcal{S}_{k,i}^{+-}(1-\mathcal{S}_{j,k}^{++}
   \mathcal{S}_{k,i}^{--})^{-1}\mathcal{S}_{j,k}^{+-},\nonumber \\
\mathcal{S}_{j,i}^{-+}=&\mathcal{S}_{j,k}^{-+}(1-\mathcal{S}_{j,k}^{++}
   \mathcal{S}_{k,i}^{--})^{-1}\mathcal{S}_{k,i}^{-+},\nonumber \\
\mathcal{S}_{j,i}^{--}=&\mathcal{S}_{j,k}^{--}
+\mathcal{S}_{j,k}^{-+}\mathcal{S}_{k,i}^{--}
   (1-\mathcal{S}_{j,k}^{++}\mathcal{S}_{k,i}^{--})^{-1}
   \mathcal{S}_{j,k}^{+-}.
\label{sec_3_20}
\end{align}
It is clear from the above that the $\mathcal{S}_{j,i}$ 
matrix is not merely a product of two matrices
$\mathcal{S}_{j,k}$ and $\mathcal{S}_{k,i}$, but rather a complicated
nonlinear composition of them.
It is important however to note that despite its evident complexity, 
such a construction of the scattering matrix
is numerically stable, as opposed to the transfer matrix method which 
fails if a system with a large number of
discontinuity points $x_{i}$ is considered. Stability of such an algorithm
has been proven in our numerical investigations by checking that the condition
(\ref{unit_5}) is satisfied with an error smaller than $10^{-14}$.
Such an accuracy can never be achieved for systems with a large
number of discontinuity points if the transfer matrix is applied.

\section{Photo-emission}

In our model investigations, we concentrate on some essential
features of the solid-vacuum interface, as exemplified by the
Sommerfeld model, in which the band structure is neglected.
This simplification allows us to consider a quite general form
of the laser field. To be more specific, the solid surface is described
by a continuous step potential
\begin{equation}
V(x)=V_0g(x/w_0),
\label{ph1}
\end{equation}
with
\begin{equation}
g(x)=1/(1+\mathrm{e}^{-x}).
\label{ph2}
\end{equation}
The parameter $w_0$ determines the skin depth of a surface. For $w_0=0$, the surface potential represents the step function, commonly used in the Sommerfeld model. In our illustrations, we put $w_0=5$. We apply our theory to the gold surface and assume that the electron effective mass is close to
the free electron mass. The work function and the Fermi energy for the gold metal are equal to 5.1 and
5.53 eV, respectively. This means that the constant $V_0$ above (as the sum of the work function and the Fermi energy) equals 10.63 eV. 

The surface potential described above can be generalized further to meet conditions suitable for other solids. In particular, one can take into account the space-dependent effective mass of electrons in semiconductor heterostructures or metals with effective masses different from the free
electron mass.

On the other hand, the form of the laser field is assumed to depend on both space and time coordinates. Since, for laser pulses of duration $\sim 30$ fs and the 800 nm wavelength, the monochromatic approximation works well, we therefore adopt the following form for the laser electric field:
\begin{equation}
\mathcal{E}(x,t)=\mathcal{E}_0(x)\sin(\omega t)=\mathcal{E}_0f_L(x)\bigl(1+\epsilon f_P(x)\bigr)\sin(\omega t),
\label{ph3}
\end{equation}
where
\begin{equation}
f_L(x)=g(x/\zeta_L-a_L)g(b_L-x/\mu_L),
\label{ph4}
\end{equation}
and similarly
\begin{equation}
f_P(x)=g(x/\zeta_P-a_P)g(b_P-x/\mu_P).
\label{ph5}
\end{equation}
The parameter $\epsilon$ defines the plasmon-enhanced part of the laser field.
For the incident laser beam, we choose the Ti:sapphire laser beam of frequency $\omega=1.5498$ eV ($\lambda=800$ nm). This means that inside the solid the laser field intensity averaged over the time period decays exponentially,
\begin{equation}
I(x)\sim \mathrm{e}^{2x/\zeta_L}.
\label{ph6}
\end{equation}
On the other hand, in vacuum, it stays constant close to the surface, and then again decays exponentially. In this way, we can mimic a real physical situation in which the radiation-filled space is finite. In our illustrations, we take $\zeta_L=40$, which means that the penetration depth of the laser field intensity equals $\zeta_L/2=20$. The parameter $a_L\zeta_L$ describes the distance in a solid at which the intensity is not reduced substantially. On the other hand, $b_L\mu_L$ corresponds to the laser focus diameter in vacuum, whereas $\mu_L$ alone determines the intensity reduction rate outside the focus. Similar parameters with the subscript $P$ refer to the plasmon-enhanced part of the laser field. The remaining parameters have been chosen as follows: $a_L=3$, $b_L=20$, $\mu_L=100$, $a_P=1$, $\zeta_P=8$, $b_P=4$, $\mu_P=20$, and $\epsilon=0,\ldots, 5$. All dimensional parameters are in atomic units.

\begin{figure}
\includegraphics[width=7cm]{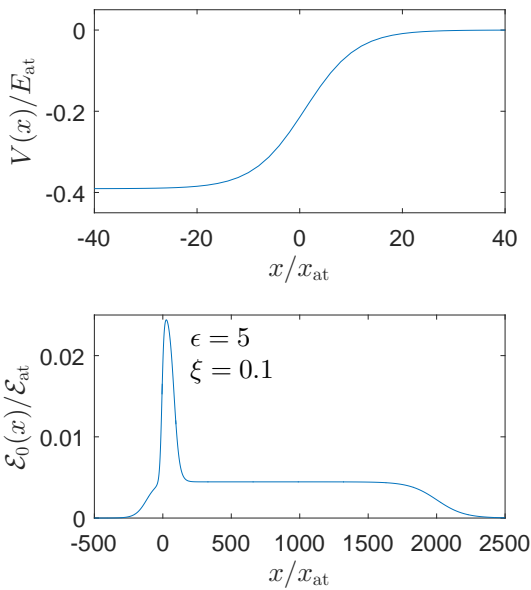}
\caption{(Color online) The continuous step potential (upper frame) and the space-dependent electric field amplitude of the laser field (lower frame). The atomic units of length, energy and the electric field strength are $x_{\mathrm{at}}\approx 0.053\,\mathrm{nm}$, $E_{\mathrm{at}}\approx 27.21\,\mathrm{eV}$ and $\mathcal{E}_{\mathrm{at}}\approx 5.14\times 10^{11}\,\mathrm{V/m}$, respectively.
}
\label{syst}
\end{figure}

In our discussions presented below, the laser field intensity is characterized by the dimensionless parameter $\xi=U_p/\omega$, where $U_p=\mathcal{E}_0^2/(4\omega^2)$ is the ponderomotive energy of electrons in the monochromatic electromagnetic plane wave of frequency $\omega$; hence $\mathcal{E}_0=2\omega\sqrt{\omega\xi}$. In Fig.~\ref{syst} we draw the space-dependence of the continuous step potential $V(x)$ and the electric field amplitude $\mathcal{E}_0(x)$ for $\epsilon=5$ and $\xi=0.1$.

\begin{figure}
\includegraphics[width=7cm]{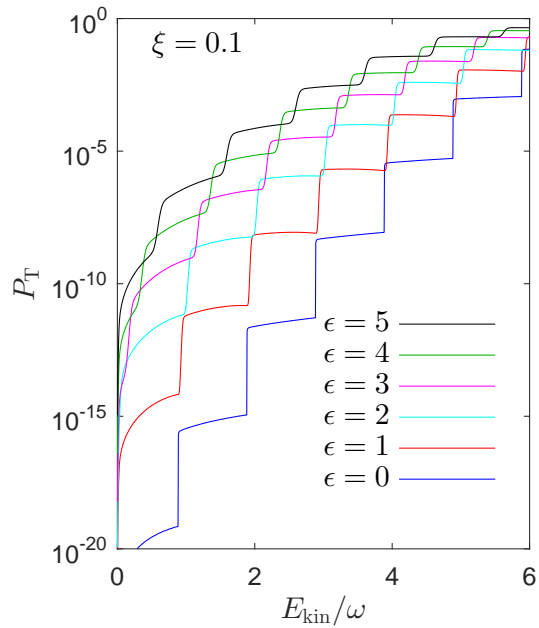}
\caption{(Color online) Total photo-emission probabilities as functions of the kinetic energy of electrons for $\xi=0.1$ and for six values of $\epsilon$. 
}
\label{trans10}
\end{figure}

The total photo-emission probability is equal to
\begin{equation}
P_{\mathrm{T}}=\sum_{N\geqslant N_{\mathrm{tr}}}^{}\frac{m_{0}q_{N}}{m_{L}p_{0}}
   {\arrowvert \textsc{T}_{N}\arrowvert}^{2}.
\label{ph7}
\end{equation}
We plot it in Fig.~\ref{trans10} as a function of the electron kinetic energy for $\xi=0.1$ and for six values of $\epsilon$. We clearly see the multi-photon structure in this distribution, i.e., the total probability jumps sometimes by a few orders of magnitude if the smaller number of laser photons is sufficient for photo-emission. As expected, the plasmon effect usually increases the photo-emission probability. Moreover, the energy of the multi-photon channel opening increases with increasing $\epsilon$, which is due to the increase of the space-dependent ponderomotive energy of the laser field. The significance of this effect for the tunneling phenomena is going to be discussed in due course.

\section{Conclusions}

As mentioned above, our algorithm is convergent provided that 
a sufficient number of
discretization points is introduced. For systems
considered here, this number should not be smaller
than 100. If the laser field is very weak, this does not
create significant numerical problems, except that
 calculations become longer. However, when the
laser field is sufficiently intense, the algorithm
based on the transfer matrix is unstable. This instability is due
to the existence of closed channels, which introduce
into numerical calculations very small and very large
numbers at the same time. Augmenting precisions significantly slows 
down the calculation and does not diminish the problem.  
We have found that it is
possible to make this algorithm numerically stable by just
applying nonlinear matrix transformations, 
without introducing higher precisions.

Illustrations presented in this paper show that photo-emission of electrons can be changed significantly by applied nonperturbative oscillating in time and space-dependent electric field. The efficiency of the algorithm presented in this contribution opens up the possibility of investigating surface phenomena in the presence of more realistic laser pulses that gradually decrease within solids
and extend on a mesoscopic scale in vacuum.

\section*{Acknowledgements}

This work is supported by the Polish National Science Center (NCN) under Grant No. 2012/05/B/ST2/02547.

\end{document}